\documentclass[showpacs,twocolumn,floatfix,superscriptaddress]{revtex4}
\usepackage{graphicx}

\newcommand{\bq}{\begin{equation}}
\newcommand{\ee}{\end{equation}}
\newcommand{\fr}[2]{\frac{#1}{#2}}
\newcommand{\eps}{\varepsilon}

\begin{document}

\title{ Minimal realization of the Orbital Kondo effect in a Quantum Dot
with two Leads

}

\author{P.G. Silvestrov}
\affiliation{Theoretische Physik III, Ruhr-Universit{\"a}t Bochum,
44780 Bochum, Germany}
\author{Y. Imry}
\affiliation{Department of Condensed-Matter Physics, Weizmann
Institute of Science, Rehovot 76100, Israel}

\date{\today }

\begin{abstract}

We demonstrate theoretically how the Kondo effect may be observed
in the transport of spinless electrons through a quantum dot. The
role of conduction electron spin is played by a lead index. The
Kondo effect takes place if there are two close levels in the dot
populated by a single electron. For temperatures exceeding the
Kondo temperature $T\gg T_K$, the conductance is maximal if the
levels are exactly degenerate. However, at zero temperature, the
conductance is zero at the $SU(2)$ symmetric point but reaches the
unitary limit $G = e^2/h$ for some finite value of the level
splitting $\Delta\eps\sim T_K$. Introducing the spin-$1/2$ for
electrons  and having two degenerate orbital levels in the dot
allows to observe the $SU(4)$-Kondo effect in a single dot coupled
to two leads.

\end{abstract}

 \maketitle

\section{Introduction}

For several decades the Kondo~effect~\cite{Kondo64} has been one
of the most attractive and widely investigated many body phenomena
in solid state physics~\cite{Hewson}. Experimental observation of
the Kondo effect in quantum dots~(QD)~\cite{exp}, also predicted
almost two decades ago~\cite{GlazmanRaikhNgLee}, started a new
wave of interest in the field. The minimal realization of the
Kondo effect requires a single electron channel with spin, coupled
to the impurity spin. This may not necessarily be a true spin, but
also some internal degree of freedom leading to the orbital Kondo
effect. Since in transport experiments the QD is coupled to two
leads, the questions why the spin is necessary for the development
of Kondo correlations and why the two leads cannot cause an
orbital Kondo effect arises. For a single active discrete level in
the QD, one lead may be effectively decoupled by a simple unitary
transformation~\cite{GlazmanRaikhNgLee}, eliminating any spinless
Kondo behavior. This could cause a belief that there is no Kondo
effect in a single dot without spin. However, as was already shown
in Ref.~\cite{Boese01}, having several orbital levels in the dot
will cause no decoupled lead to exist and enable the Kondo
correlations without spin. As we will show in this paper, the
transport through such a QD has a very peculiar form, where both
the temperature and gate-voltage dependencies are nonmonotonous.
At zero temperature the conductance vanishes for exactly
degenerate levels $\Delta\eps =0$ when the (pseudo)spin is
completely screened, while it reaches the unitary limit at finite
level splitting $\Delta\eps\approx 2.43T_K$, where the screening
is only partial. (Similar behavior for a related double dot system
was found recently in Ref.~\cite{Meden06} using the functional
renormalization-group approach.) On the other hand, at high
temperatures the conductance is maximal at the degeneracy point.
Such rich transport properties in the Kondo regime clearly call
for an experimental verification.

If the spin degeneracy is present in addition to the two close
orbital levels, the $SU(4)$-Kondo effect develops in a single QD
with two single-channel leads. Again the conductance of a QD in
the vicinity of $SU(4)$ degenerate point has a peculiar
gate-voltage and/or temperature behavior. This behavior may be
used for spintronics applications, as suggested in
Ref.~\cite{Halperin03} for double-dot devices.

Even for two degenerate levels in the closed dot, the Kondo
correlations do not always develop. Virtual electron jumps from
the QD to the leads and back both shift the single electron
energies in the dot~\cite{Sil00,Sil02} and induce an effective
coupling between localized states, which lifts out the degeneracy.
Consequently, in order to reach the $SU(2)/SU(4)$ symmetric point,
one needs to tune two parameters, the level splitting and the
non-diagonal matrix element, which in the following we will
parameterize by two components of the {\it effective} magnetic
field, $B_z$ and $B_x$. Such  two-parameter tuning in a single QD
may still be easier experimentally than manipulating the double
QD. Concurrently, the effective magnetic field dependence of the
conductance may serve as an indication of the orbital Kondo effect
in a single QD.

We first consider the case of spinless electrons, where analytical
results are easily available for both $T\gg T_K$ and $T\ll T_K$.
This case may be achieved experimentally, for example, by lifting
the spindegeneracy with an external magnetic field (see below).
The $SU(4)$-Kondo effect will be considered in the concluding part
of the paper.

Experimentally both orbital and $SU(4)$-Kondo effects were
observed recently in carbon nanotube QDs~\cite{Jarillio05}. The
effect in this case originates from the extra valley degeneracy of
electronic states in carbon nanotubes, which thus takes place both
in the QD and in the leads. Several theoretical papers addressed
the possibility of having orbital and $SU(4)$-Kondo effect in
nanotubes, double QDs and QDs with several
leads~\cite{Halperin03,Eto05,FSE}. Reference~\cite{Boese01} first
predicted the possibility of orbital Kondo effect in the case of
QD coupled to two leads. In both carbon nanotubes and double QDs
the orbital and/or $SU(4)$-Kondo effect should be seen in the
whole Coulomb blockade valley separating two charging resonances.
The emphasis of the current research is on surprisingly rich
conductance behavior of the QD with two leads in the Kondo regime,
which develops in a narrow region in the parameter space near the
degeneracy point.

\section{Effective intradot Hamiltonian}

The Hamiltonian of a quantum dot with two close levels ($j=1,2$)
coupled to two leads ($L,R$) has the form
 \begin{eqnarray}\label{Ham}
&& \! \! \! \! \! \! \! \! \! \! \! \! H=\sum_{ik}\eps_k
c_{ki}^{\dagger}c_{ki}+\sum_j \eps^d_j d_j^\dagger
d_j+\fr{U}{2}\sum_{ij}d_i^\dagger d_i d_j^\dagger d_j\\
&& \ \ \ \ \ \ \ \ +\sum_{ijk}(t_{ji} c_{ki}^{\dagger} d_j+{\rm
H.c.}).
 \nonumber
 \end{eqnarray}
Here, $d_j$ ($c_{ki}$) annihilates the electron at the level $j$
in the dot (the one with momentum $k$ in the lead $i$);
$\eps^d_{j}$ ($\eps_k$) are the electron energies in the dot
(lead); $U$ is the charging energy of the dot and $t_{ji}$ are the
dot-lead coupling matrix elements. We take all $t_{ji}$ to be
real. (This requires that while the Zeeman splitting of spin-up
and -down states exceeds the energy difference of two close
levels, the magnetic field is parallel to the plane of the dot or
small enough to not affect the orbital states in the dot). When
the Fermi energy $\eps_F$ in the leads is chosen as $\eps^d_j<
\eps_F< \eps^d_j +U$, the dot is charged by one electron.

In the second order of perturbation theory two processes lead to
the renormalization of the single-particle interdot Hamiltonian:
Electron jumps from the dot to the lead or  vice versa, both
followed by the return of an electron. The corresponding matrix
elements calculated in~\cite{Sil02} are (see a similar calculation
for the Anderson impurity model in Ref.~\cite{Haldane78}):
 \begin{eqnarray}\label{HamEff}
V_{ij}=\eps^d_i\delta_{ij}-\fr{\Gamma_{ij}}{2\pi}\ln
\fr{E-{\eps_{F}}+U}{|E-{\eps_{F}}|} .
 \end{eqnarray}
Here $\Gamma_{ij} = 2\pi\nu\sum_{l=L,R} t_{il} t_{jl}$, $i,j$ label
the level indices $1$ or $2$ and $E=(V_{11}+V_{22})/2$. For
simplicity we assume the same density of states $\nu =dn/d\eps$ in
the two leads. The Hamiltonian of the isolated dot now takes the
form
 \bq\label{HamDot}
H_d=E\sum_i d_i^\dagger d_i+\fr{U}{2}\sum_{ij}d_i^\dagger d_i
d_{j}^\dagger d_{j} +\vec{B}\vec{S},
 \ee
where $\vec{S}=\sum_{ij}d_i^\dagger \vec{\sigma}_{ij} d_{j}/2$ and
the only nonzero components of the effective magnetic field are
$B_x=2V_{12}$ and $B_z=V_{11}-V_{22}$. We assume that
$|\vec{B}|\ll U$, while $\eps_F-E\sim U$.

The QD Hamiltonian, Eq.~(\ref{Ham}), formally coincides with the
Hamiltonian of the double-dot system tunnelcoupled to two shared
leads, investigated in Ref.~\cite{Meden06}. Consequently our
result for the conductance in the spinless case found in Sec.~V
for strongly developed Kondo correlations ($U\gg\Gamma,T=0$)
closely resembles the correlation-induced resonances found in that
paper for intermediate values of~$U/\Gamma$.

\section{The Kondo Hamiltonian}

A convenient way to define the pseudospin states is to perform
unitary transformations of both dot and lead states to diagonalize
the matrix $U_l T U_d^\dagger$, where the matrix $T$ is formed by
the tunnelling matrix elements $t_{ij}$ [Eq.~(\ref{Ham})] and the
unitary matrices $U_l(U_d)$ operate in the lead-index(dot-level)
space. Fortunately, in the case of interest to us, this procedure
is simplified. In order to have strong Kondo correlations, the
effective Zeeman splitting in the dot should be small compared to,
e.g., the widths of the levels (we require both components
$B_x\sim B_z$ to be small, not the vanishing of one of them).
This, in particular, implies $|\Gamma_{12}|\ll \max
(\Gamma_{11},\Gamma_{22})$. Since $\Gamma_{12}\sim\sum t_{1i}
t_{2i}$, a simple rotation in the lead-index space (angle $\alpha,
\tan\alpha=t_{1R}/t_{1L}\approx -t_{2L}/t_{2R}$) allows to define
the mixed lead states, each coupled to one of the levels in the
dot with the tunnelling matrix elements
 \bq\label{t12}
t_1=\sqrt{t_{1L}^2+t_{1R}^2} \ , \ t_2=\sqrt{t_{2L}^2+t_{2R}^2}.
 \ee

The standard Schrieffer-Wolf transformation leads us now to the
Kondo-type Hamiltonian
 \bq\label{HKondo}
H_K= \sum_{ki} \eps_k c^{\dagger}_{ki} c_{ki}
 + {J}({s}_xS_x+{s}_yS_y)
+J_z{s}_zS_z +\vec{B}\vec{S},
 \ee
where ${\vec s}=\sum_{k qij}c^{\dagger}_{ki} {\vec
\sigma}_{ij}c_{qj}/2$, and we dropped out irrelevant terms
describing electron scattering without spin-spin interaction.

\begin{figure}
\includegraphics[width=7.5cm]{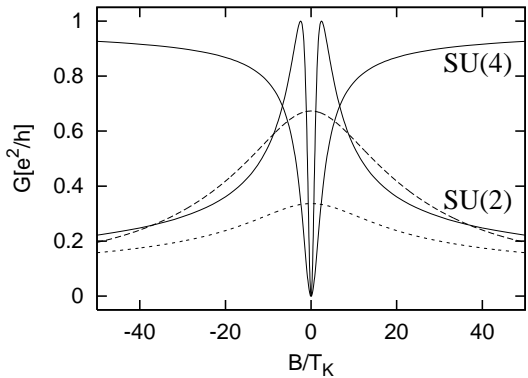}
\caption{ The conductance of a quantum
dot~(\ref{Ham},\ref{HKondoK}) as a function of $B/T_K$ for $\sin
2\phi=1$. Lower short-dashed line: high temperature conductance,
$T=15T_K$, but with only non-spin-flip processes taken into
account, Eq.~(\ref{21}). The cut off in Eq.~(\ref{KKK}) is chosen
in a form $D=\sqrt{B^2+T^2}$. Long-dashed line: the same
$T=15T_K$, but with the spin-flip current added, Eq.~(\ref{23}).
The conductance is doubled at $B\ll T$. Lower solid
line~[$SU(2)$]: exact Bethe-Ansatz at
$T=0$~[Eq.~(\ref{conducT0})]. The conductance vanishes at $B=0$
and reaches the unitary limit, $G=e^2/h$, at $B=2.43T_K$. Upper
solid line: schematic conductance (divided by $2$) in the case of
two orbital levels with spin~[Eq.~(\ref{conducT04})]. The
conductance vanishes at the $SU(4)$ symmetric point, $B=0$, and
reaches the unitary limit, $G=2e^2/h$, at $|B|\gg T_K$. The
conductance at high temperatures in the case with spin is
qualitatively similar to the conductance for spinless electrons,
having a pronounced maximum at $B=0$~[Eq.~(\ref{GtempSU4})].}
\end{figure}

The bare exchange spin-spin scattering amplitudes
 \bq\label{Jbare}
J_{z0}=\fr{(t_1^2+t_2^2)U}{|E-\eps_F|(E-\eps_F+U)} \ , \
J_0=\fr{2t_1 t_2}{t_1^2+t_2^2}J_{z0}
 \ee
are renormalized by integrating out the electron states in the
leads with $|\eps_k -\eps_F|>D$. The corresponding
renormalization-group equations have the form
 \begin{eqnarray}
J_z'=-J^2 \ , \ 
J'=-JJ_z.
 \end{eqnarray}
Here the derivative $A'=dA/d(\nu \ln D)$. The two differential
equations have an "integral of motion",
$\lambda=\sqrt{J_z^2-J^2}={\rm constant}$. A simple calculation now
gives
 \bq\label{KKK}
J_z=\lambda{\rm coth}\left[\lambda\nu\ln\fr{D}{T_K}\right] \ , \
\fr{T_K}{U}=\exp\left[ \fr{-1}{\lambda\nu}| \ln\fr{t_1}{t_2} |
\right],
 \ee
where $T_K$ is the Kondo temperature.

Once the renormalization of the spin-spin interaction becomes
sufficient, $J(D)\gg J_0$, one may neglect the difference between
$J$ and $J_z$. Then, the scattering on the impurity become $SU(2)$
invariant and we may again perform a rotation in the pseudospin
space leading to
 \bq\label{HKondoK}
 H_K= \sum_{ks} \eps_{k} c^{\dagger}_{ks} c_{ks}+J\vec{s}\vec{S}+
 BS_z,
 \ee
where $B=\sqrt{B_z^2+B_x^2}$. The lead spin states are
 \bq\label{phi+alpha}
c_{k\uparrow} ={\rm c} \, c_{kL} +{\rm s} \, c_{kR} \ , \
c_{k\downarrow} =-{\rm s} \, c_{kL} +{\rm c} \, c_{kR},
 \ee
where ${\rm c}=\cos\phi$, ${\rm s}=\sin\phi$ and the angle $\phi$
includes two contributions, the rotation of lead states which
diagonalizes the matrix $T=(t_{ij})$ and the rotation of both the
lead-electron and the dot spins in order to align the $z$~axis
with the direction of effective magnetic field
 \bq\label{phi}
\phi=\fr{1}{2}\arctan \fr{B_x}{B_z} +\arctan \fr{t_{1R}}{t_{1L}} .
 \ee

\section{The linear conductance}

First we consider the situation where the Kondo correlations are
developed but have not reached the unitary limit, $\nu J_0\ll \nu
J\ll 1$. Let also the temperature $T$ and the voltage
$eV=\eps_{F_L}-\eps_{F_R}$ be small compared to the effective
Zeeman splitting in Eq.~(\ref{HKondoK}), $B\gg T,eV$. Then the
spin projection of the QD is always $S_z=-1/2$ with no
fluctuations. The transport is now entirely due to the term
$Js_zS_z\rightarrow \fr{J}{4}\sin 2\phi\sum_k c^\dagger_{kR}
c_{kL}$ in Eq.~(\ref{HKondoK}), leading to the conductance
 \bq\label{21}
G=\fr{e^2}{h}\left[\sin 2\phi \fr{\pi \nu J(B)}{2} \right]^2 =
\fr{e^2}{h}\left[ \fr{B_z}{B} \fr{\pi \nu J(B)}{2}\right]^2.
 \ee
In the last formula we substituted the angle
$\phi$~[Eq.~(\ref{phi})] for symmetric dot-to-lead coupling,
$t_{1R}=t_{1L}$. A different choice of the ratio $t_{1R}/t_{1L}$
would lead to a simple rotation on the $(B_x,B_z)$ plane. The
Kondo correlations enter Eq.~(\ref{21}) through the running
coupling constant~$J(B)$.

Slightly more complicated is the case of a finite temperature
$T\gg T_K$. In the spirit of Ref.~\cite{Beenak92} we may obtain
the probability to find the dot in one of the spin states
$n_\uparrow$ and $n_\downarrow$ via the stationary solution of
corresponding master equation with small applied bias
$\eps_{F_L}=\eps_{F_R}+eV$,
 \bq\label{occupations}
\fr{n_\downarrow}{n_\uparrow}=\exp(B/T)\left[ 1- \fr{eV}{T}\cos
2\phi\right] .
 \ee
Then, we obtain the conductance, which includes both non-spin-flip
and spin-flip currents, as follows:
 \bq\label{23}
G=\fr{e^2}{h}\left[ \sin 2\phi \fr{\pi \nu J}{2} \right]^2 \Big(
1+\fr{B/T}{\sinh (B/T)} \Big).
 \ee
Here the running coupling constant $J$~[Eq.~(\ref{KKK})] may be
calculated, e.g., at the average $D=\sqrt{B^2+T^2}$. Details of
the derivation of Eqs.~(\ref{occupations}) and (\ref{23}) are
given in the Appendix.

Theoretically the $\sim (J/J_0)^2$ increase of the conductance due
to the Kondo effect may become very large. However, since the spin
scattering constant $J$ depends only logarithmically  on the
cutoff $D$~[Eq.~(\ref{KKK})], the more pronounced in the
experiment may be the enhancement due to the last factor in
Eq.~(\ref{23}), which accounts for the spin-flip processes near
the $SU(2)$ symmetric point. The non-spin-flip and spin-flip
contributions to the conductance may be measured separately in a
QD embedded into the Aharonov-Bohm
interferometer~\cite{Schuster97}

Conductance as a function of $B$ in various regimes as well as the
contour plot of the conductance on the $B_z,B_x$ plane are shown
in Figs.~1 and ~2.

\section{Exact $S$-matrix}\label{exact}

For  $T\ll T_K$, the electron is only elastically scattered on the
Kondo impurity and the conductance is determined by a
single-particle $S$-matrix~\cite{Noziers,Pustil01}. In the usual
case of electron with spin-$1/2$ and single-level QD the
$S$-matrix connects four incoming and four outgoing waves,
corresponding to two "spin orientations" in the two leads.
However, only two effective channels acquire a nontrivial
scattering phase, while the other two are completely
decoupled~\cite{GlazmanRaikhNgLee}. In our setup there are only
two channels  with nontrivial phase behavior, corresponding to the
two pseudo-spin directions, and the $S$-matrix in terms of these
effective channels entering the Kondo Hamiltonian
Eq.~(\ref{HKondoK}) has the form~\cite{endnote}
 \bq\label{SSU2}
S= \left(\begin{array}{cc}
e^{2i\delta}&0\\
0&e^{-2i\delta}
\end{array}\right),
 \ee
where the phase $\delta=\delta(B/T_K)$ increases from $0$ to $\pi$
when $B$ changes from $+\infty$ to $-\infty$ and $\delta(0)=\pi/2$.
In order to find the $S$-matrix in the original $L-R$ basis we
perform a rotation (${\rm c}=\cos\phi, {\rm s}=\sin\phi$)
 \bq\label{rotat}
S^{(LR)}= \left(\begin{array}{cc}
{\rm c}&-{\rm s}\\
{\rm s}&{\rm c}
\end{array}\right)
\left(\begin{array}{cc}
e^{2i\delta}&0\\
0&e^{-2i\delta}
\end{array}\right)
\left(\begin{array}{cc}
{\rm c}&{\rm s}\\
-{\rm s}&{\rm c}
\end{array}\right).
 \ee
The nondiagonal matrix element  $S^{(LR)}_{12}=i\sin 2\phi\sin
2\delta$ now determines the conductance
 \bq\label{conducT0}
G=\fr{e^2}{h}\sin^2 2\phi \sin^2 2\delta .
 \ee
The explicit dependence of the phase $\delta(B/T_K)$ is found from
the Bethe ansatz solution~\cite{WiegmannTsvelick}. When
approaching the $SU(2)$ symmetric point the conductance first
increases, reaching the unitary limit ($\delta=\pi/4, 3\pi/4$) at
$B=\pm 2.43T_K$. It vanishes, however, at the symmetric point,
$\delta(0)=\pi/2$~\cite{endPustil01}.

\begin{figure}
\includegraphics[width=7.5cm]{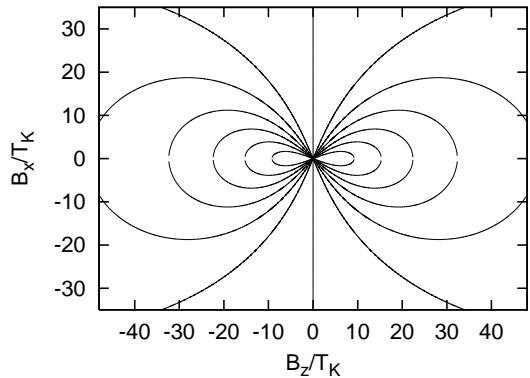}
\caption{Contour plot of the conductance on the $B_z,B_x$ plane at
$T=15T_K$, Eq.~(\ref{23}). The conductance is maximal at
$\vec{B}=0$, but the peak has zero width in the $B_x$ direction.
At zero temperature a valley with conductance $G(B_z=0)=0$
develops along the $B_x$ axis, separating two peaks with
$G=e^2/h$. }
\end{figure}

\section{$SU(4)$-Kondo Effect}

Taking into account the true electron spin is achieved in
Eqs.~(\ref{Ham}) and (\ref{HamDot}) by simply adding an extra
index to all creation and annihilation operators. The bare Kondo
Hamiltonian [analog of Eq.~(\ref{HKondo})] will now contain a
number of new operators describing both spin, isospin, and mixed
spin-pseudo-spin scattering \cite{Halperin03}. Fortunately, the
renormalization-group flow of the corresponding constants at low
temperatures (and small level splitting in the dot) results in a
simple $SU(4)$ symmetric Hamiltonian, known as a
Coqblin-Schrieffer model~\cite{Coqb69},
 \bq\label{Coqblin}
H_{\rm eff} =\tilde{J} \!\!\! \sum_{\alpha,\beta =1,...,4} \!\!\!
[\psi_\alpha^\dagger\psi_\beta |\beta\rangle\langle\alpha |
-\fr{1}{4}\psi_\alpha^\dagger\psi_\alpha |\beta\rangle\langle\beta
|]+BT_z .
 \ee
Here the isomagnetic field $B$ is the same as in
eq.~(\ref{HKondoK}), but we have replaced $S_z$ by ${T_z}$ to
emphasize that this is an isospin operator, $|\alpha\rangle$'s
denote the four states of quantum dot, and $\psi_\alpha= \sum_k
c_{k\alpha}$.

Integrating out the lead states  with $|\eps-\eps_F|>D$, increases
the coupling constant $\tilde{J}$ like $\tilde{J}\sim
1/\ln(D/T_K)$. This renormalization of Eq.~(\ref{Coqblin})
continues until $D\approx B$. At $D<B$ the $SU(4)$ symmetry of the
Hamiltonian is broken to $SU(2)$ (single level with spin $s=1/2$).
The $SU(2)$ coupling constant is further (logarithmically)
renormalized with decreasing $D$, but the corresponding increase
of the conductance is slower than for the $SU(4)$ symmetric case.
This effect was interpreted in Ref.~\cite{Eto05} as the increase
of the Kondo temperature close to the degeneracy point (as done
before for the case of singlet to triplet
transition~\cite{EtoNaz00}). However, these $SU(4)$ vs $SU(2)$
renormalization group effects change $\tilde{J}$ only through
logarithmic corrections. Thus in a large range of variation of
parameters the most pronounced effect will be the conductance
increase at $B\le T$ due to spin-flip processes. A straightforward
calculation shows that Eq.~(\ref{occupations}) for intra-dot
levels occupation remains valid in the presence of the true spin
and the conductance is given by
 \bq\label{GtempSU4}
G=\fr{2e^2}{h}(\pi\nu\tilde{J}\sin 2\phi)^2\Big( 1+\fr{B/T}{\sinh
(B/T)} \Big).
 \ee

At zero temperature, electron scattering is described by a
$4\times 4$ $S$-matrix, which is diagonal in the basis
Eq.~(\ref{phi+alpha})
 \bq\label{Sdiag}
S= {\rm diag}( e^{2i\delta_1}, e^{2i\delta_1}, e^{2i\delta_2},
e^{2i\delta_2}) .
 \ee
When $B$ changes from $+\infty$ to $-\infty$, the first phase
decreases from $\delta_1=\pi/2$ to $\delta_1=0$, while the second
increases from $\delta_2=0$ to $\delta_2=\pi/2$~\cite{endnote2}.
The $SU(4)$ symmetric point $B=0$ corresponds to
$\delta_1=\delta_2=\pi/4$ (see Ref.~\cite{Halperin03} for a NRG
calculation of phases). Transformation to the original $L-R$ basis
is done via a rotation, similar to that of Eq.~(\ref{rotat}),
separately of the pairs of states $1-3$ and $2-4$. This gives
(compare Ref.~\cite{Pustil01})
 \bq\label{conducT04}
G=\fr{2e^2}{h}\sin^2 2\phi \sin^2 (\delta_1-\delta_2) .
 \ee
As in the $SU(2)$ case [Eq.~(\ref{conducT0})], the conductance
vanishes at the $SU(4)$ symmetric point $\delta_1=\delta_2$. Away
from this point the usual spin-$1/2$ Kondo effect takes place due
to transport through the lower of the two levels in the dot. The
conductance is maximal here since $\sin^2
(\delta_1-\delta_2)\approx 1$ (unitary limit).

\section{Conclusions}

In this paper, we discussed how with suitable tuning of the dot
levels and the couplings to the two leads an orbital $SU(2)$-Kondo
effect can be realized for spinless electrons and {\em  a single
Quantum Dot}. We obtained the nontrivial transport properties of
this model. Introducing the real electron spin results in the
$SU(4)$-Kondo effect. A nonmonotonic conductance dependence on the
temperature and the effective magnetic field should allow the
experimental verification of our findings. One of our important
predictions is that the conductance, which is zero at the point of
complete screening of the dot pseudo-spin, $B,T=0$, increases all
the way up to the unitary limit if the ideal screening is
destroyed by the effective magnetic field. The question what is
the maximal value of the conductance in case of partial screening
due to finite temperature remains open. Solving this problem would
require a Numerical Renormalization-Group solution at $T$ around
$T_K$.

{\bf Acknowledgements.} We thank Yuval Oreg for helpful
discussions. This work was supported by the SFB TR 12, by the
German Federal Ministry of Education and Research (BMBF) within
the framework of the German-Israeli Project Cooperation (DIP) and
by the Israel Science Foundation (grant No. 1566/04). PGS's visit
at Weizmann was supported by the EU - Transnational Access
program, EU project RITA-CT-2003-506095. When this paper was
prepared for publication we became aware of independent
calculations on a related problem by V.~Kashcheyevs {\it et
al}~\cite{Kashcheyevs06}. After the submission we also became
aware of a preprint~\cite{HyunWoon06}, whose content partly
overlaps with our Sec.~\ref{exact}.

\appendix

\section{ Finite Temperature}

In this appendix, we present the details of derivation of
Eqs.~(\ref{occupations}),(\ref{23}), and (\ref{GtempSU4}) for
level occupations in the dot and conductance in the case of high
temperature $T\gg T_K$. Our calculation may be considered as a
generalization of the method of Ref.~\cite{Beenak92}. Specific for
our problem is that the quantum dot is out of resonance and the
elementary processes which should be balanced in the stationary
solution of the master equation are not adding(removing) an
electron to the dot but transmission of the electron through the
dot accompanied by the switch of the dot internal state. Also, in
our case, what is called the effective spin-up and spin-down
states of the leads are the quantum superpositions of the true
left and right lead states. Applying a finite bias to the QD, one
changes the Fermi energy in the left-right leads and not in the
spin up-down effective leads.

The second and third terms in the Hamiltonian Eq.~(\ref{HKondoK})
may symbolically be written as
 \begin{eqnarray}
&& \! B S_z +Js_z S_z +\fr{J}{2}(s_-S_+ + s_+S_-)=\\
&=& \! B S_z +\fr{J}{2}\big[S_z (|\!\!
\uparrow\rangle\langle\uparrow\!\! |-
|\!\!\downarrow\rangle\langle\downarrow\!\!|)+ S_+
|\!\!\downarrow\rangle\langle\uparrow\!\!| +S_-
|\!\!\uparrow\rangle\langle\downarrow\!\!|\big] \nonumber ,
 \end{eqnarray}
where, for example,
$|\!\!\uparrow\rangle=\sum_{k}c^\dagger_{k\uparrow}$. In terms of
the original left and right lead operators we may write
 \begin{eqnarray}
&&2s_z=|\!\! \uparrow\rangle\langle\uparrow\!\! |-
|\!\!\downarrow\rangle\langle\downarrow\!\!|= \\
&&({\rm c}^2-{\rm s}^2)(| L\rangle\langle L |- | R\rangle\langle
R|) + 2{\rm c}{\rm s} (|L\rangle\langle R|+ |R\rangle\langle L|),
\nonumber
 \end{eqnarray}
where ${\rm c}=\cos \phi$, ${\rm s}=\sin \phi$ (\ref{phi+alpha})
and $| L\rangle=\sum_{k}c^\dagger_{kL}, |
R\rangle=\sum_{k}c^\dagger_{kR}$. Similarly
 \begin{eqnarray}
&&s_-=|\!\!\downarrow\rangle\langle\uparrow\!\!|=\\
&&=-{\rm c}{\rm s}(| L\rangle\langle L |- | R\rangle\langle R|) -
{\rm s}^2|L\rangle\langle R|+{\rm c}^2 |R\rangle\langle
L|.\nonumber
 \end{eqnarray}
Now we may write a detailed balance equation for the probabilities
to find the dot in states up($+$) and down($-$). For example,
 \begin{eqnarray}\label{balanced}
\dot{n}_+ &=&-\dot{n}_-\propto n_-\big\{ {\rm c}^2{\rm
s}^2[\langle f^{\eps_L} g^{\eps_L+B}\rangle +\langle f^{\eps_R}
g^{\eps_R+B}\rangle]\nonumber
\\
&& + {\rm s}^4 \langle f^{\eps_R} g^{\eps_L+B}\rangle +{\rm c}^4
\langle f^{\eps_L} g^{\eps_R+B}\rangle \big\}
- \nonumber \\
&&n_+\big\{ {\rm c}^2{\rm s}^2[\langle f^{\eps_L}
g^{\eps_L-B}\rangle +\langle
f^{\eps_R} g^{\eps_R-B}\rangle ] \nonumber \\
&& +{\rm c}^4 \langle f^{\eps_R} g^{\eps_L-B}\rangle +{\rm s}^4
\langle f^{\eps_L} g^{\eps_R-B}\rangle \big\} . 
 \end{eqnarray}
Here $f$ is a Fermi function for electrons in the lead, $g=1-f$,
and we also introduced symbolic notations
 \begin{eqnarray}
&&\langle f^{\eps_1}g^{\eps_2}\rangle =\int \fr{1}{1+\exp{(x-\eps_1)/T}}\\
&& \left( 1- \fr{1}{1+\exp{(x-\eps_2)/T}}\right) \fr{dx}{T}=
\fr{\delta e^{-\delta}}{1-e^{-\delta}}, \nonumber
 \end{eqnarray}
where $\delta =(\eps_2-\eps_1)/T$.

For zero bias, $eV=0$, Eq.~(\ref{balanced}) transforms into
 \bq\label{zerobias}
\dot{n}_+\propto \fr{b}{2\sinh b/2}[n_-e^{-b/2}-n_+e^{b/2}],
 \ee
where $b=B/T$. Now let the Fermi energies in the two leads be
different
 \bq
E_{F_L}=E_{F_R}+eV .
 \ee
In this case, instead of Eq.~(\ref{zerobias}), we get
 \begin{eqnarray}
\dot{n}_+&\propto& n_- \left(1-
({\rm c}^2-{\rm s}^2)\fr{eV}{T}\fr{d}{db}\right)\fr{b}{e^b-1}-\\
&-&n_+ \left(1- ({\rm c}^2-{\rm
s}^2)\fr{eV}{T}\fr{d}{db}\right)\fr{b}{1-e^{-b}} \ .\nonumber
 \end{eqnarray}
Finally, we end up with the remarkably simple expression
 \bq\label{nnneq}
\fr{n_-}{n_+}=e^{B/T}\left[ 1- \fr{eV}{T}\cos 2\phi\right] .
 \ee
There is a nontrivial nonequilibrium distribution even for $B=0$.
This is because of the electrons are injected through the leads,
which are themselves the superpositions of pseudospin-up and -down
states.

Conductance through the QD described by the Hamiltonian
Eq.~(\ref{HKondoK}) may be split into several contributions.
First, the non-spin-flip transitions through the ground state
$n_-\rightarrow n_-$ give
 \bq\label{n36}
I_{n_-\rightarrow n_-}= \fr{e}{h} eV n_- \left[\fr{J}{2}2{\rm
cs}\right]^2 .
 \ee
The overall coefficient may always be recovered by comparison with
Eq.~(\ref{21}). Transitions (non-spin-flip) through the excited
state differ only by the factor $n_+$ instead of $n_-$. Similarly
the transitions with flipping the impurity from the ground state
$n_-\rightarrow n_+$ give
 \begin{eqnarray}\label{n37}
I_{n_-\rightarrow n_+}&=& \fr{e}{h} (B-eV)n_- \left[{J}{\rm
s}^2\right]^2\fr{e^{-b+v}}{1-e^{-b+v}}\\
&-&\fr{e}{h} (B+eV)n_- \left[{J}{\rm
c}^2\right]^2\fr{e^{-b-v}}{1-e^{-b-v}}.\nonumber
 \end{eqnarray}
Here $v=eV/T$. The first term in the r.h.s. of Eq.~(\ref{n37})
describes the charge transfer from the left to the right lead, and
the second term accounts for for the charge transfer from the
right to the left lead. Similarly the $n_+\rightarrow n_-$
contribution reads
 \begin{eqnarray}\label{n38}
I_{n_+\rightarrow n_-}&=& \fr{e}{h} (B+eV)n_+ \left[{J}{\rm
c}^2\right]^2\fr{1}{1-e^{-b-v}}\\
&-&\fr{e}{h} (B-eV)n_+ \left[{J}{\rm
s}^2\right]^2\fr{1}{1-e^{-b+v}}.\nonumber
 \end{eqnarray}
Straightforward calculation now gives Eq.~(\ref{23}).

To make explicit the analogy between the $SU(4)$ and $SU(2)$ cases
we rewrite the term in the Hamiltonian Eq.~(\ref{HKondoK})
describing the spin-dependent scattering in the notations used in
Eq.~(\ref{Coqblin}),
 \bq\label{SU(2)}
J\vec{S}\vec{s}=\fr{J}{2}\sum_{\alpha\beta}|\alpha\rangle\langle\beta|
\ \psi_\beta^\dagger\psi_\alpha -\fr{J}{4}\sum_\alpha
|\alpha\rangle\langle\alpha| \sum_\beta
\psi_\beta^\dagger\psi_\beta .
 \ee
The only difference between the spin scattering in the $SU(4)$ and
$SU(2)$ forms is now in the relative amplitude of potential
scattering $\sum |\alpha\rangle\langle\alpha| \sum
\psi_\beta^\dagger\psi_\beta$. The potential scattering, however,
does not lead to any spin-flip processes and, in our geometry,
even does not contribute to any current at all. Indeed, in the
potential scattering term, one may always perform individual
unitary rotations of lead and dot operators so that each lead will
be connected to its own level. The electron in this case may
oscillate between lead and connected level, but will never be
transferred from one lead to another.

Equation~(\ref{balanced}) is still valid in the case of $SU(4)$
Hamiltonian Eq.~(\ref{Coqblin}). Simply now $n_+$ and $n_-$ are
the probabilities to find the dot in the pseudo-spin up or down
state with any orientation of the usual spin. Consequently, the
occupation ratio $n_-/n_+$ is still given by Eq.~(\ref{nnneq}),
only now $n_++n_- =1/2$.

In the calculation of the conductance one should take into account
that since potential scattering does not contribute to the
conductance and since the flip of usual spin does not cost any
energy, the contributions to conductance due to processes with and
without flip of spin coincide. Thus the calculation become almost
trivial. First, one replaces $J$ by $2\tilde{J}$ in
Eqs.~(\ref{21}), and (\ref{23}) due to a different definition of
the strength of Kondo interaction in Eqs.~(\ref{Coqblin}) and
(\ref{SU(2)}). Second, the conductance is doubled due to the
processes with flipping of the usual spin. Thus, we obtain
Eq.~(\ref{GtempSU4}).


\begin{thebibliography}{99}
\bibitem{endnote} In the case of a single level in the dot and an electron
with spin one should write a separate $S$-matrix for each spin
orientation ${\rm diag} (S)=(e^{\pm 2i\delta}, 1)$. A calculation
similar to Eq.~(\ref{rotat}) now gives a conductance different
from Eq.~(\ref{conducT0}) by an overall factor of 2 and a
replacement $2\delta\rightarrow \delta$
(Ref.~\cite{GlazmanRaikhNgLee}).
\bibitem{endnote2}
The phases $\pm\pi/2$ are physically indistinguishable in eqs.
(\ref{Sdiag}) and (\ref{SSU2}). This makes these two equations
consistent  at the $B\rightarrow \pm \infty$ limit.
\end{thebibliography}
\end{document}